\documentclass[twocolumn,english,aps,prd,fleqn,floatfix]{revtex4}
\usepackage{epsf,amsmath}
\usepackage{graphicx}
\usepackage{rotating}

\usepackage[T1]{fontenc}
\usepackage[latin1]{inputenc}

\makeatletter
\usepackage{babel}
\makeatother
\begin{document}

\preprint{SI-TH-2000}

\title{Antiparticle to Particle Production Ratios 
in hadron--hadron and d--Au 
collisions \\ in the \textsc{DPMJET}-III Monte Carlo}

\author{F.\:W.\:Bopp} 
\email{bopp@physik.uni-siegen.de}
\author{J.\:Ranft}
\email{Johannes.Ranft@cern.ch}
\affiliation{Fachbereich Physik, Universit\"at Siegen, D--57068 Siegen,
Germany}
\author{R.\:Engel}
\email{Ralph.Engel@ik.fzk.de}
\affiliation{Forschungszentrum Karlsruhe, Institut f\"ur Kernphysik,
Postfach
3640, D--76021 Karlsruhe}
\author{S.\:Roesler}
\email{Stefan.Roesler@cern.ch}
\affiliation{CERN, Geneva, Switzerland}

\begin{abstract}

To understand baryon stopping we analyse new RHIC and Fermilab data 
within the framework of the multichain Monte Carlo  
\textsc{DPMJET}--III.   The present  consideration is restricted to 
to hadron--hadron and  d--Au 
collisions.
\end{abstract}
\maketitle

%
\newbox\hdbox%
\newcount\hdrows%
\newcount\multispancount%
\newcount\ncase%
\newcount\ncols
\newcount\nrows%
\newcount\nspan%
\newcount\ntemp%
\newdimen\hdsize%
\newdimen\newhdsize%
\newdimen\parasize%
\newdimen\spreadwidth%
\newdimen\thicksize%
\newdimen\thinsize%
\newdimen\tablewidth%
\newif\ifcentertables%
\newif\ifendsize%
\newif\iffirstrow%
\newif\iftableinfo%
\newtoks\dbt%
\newtoks\hdtks%
\newtoks\savetks%
\newtoks\tableLETtokens%
\newtoks\tabletokens%
\newtoks\widthspec%
%
%
\immediate\write15{%
CP SMSG GJMSINK TEXTABLE --> TABLE MACROS V. 851121 JOB = \jobname%
}%
%
%
\tableinfotrue%
\catcode`\@=11
\def\out#1{\immediate\write16{#1}}
%
%
\def\tstrut{\vrule height3.1ex depth1.2ex width0pt}%
\def\and{\char`\&}
\def\tablerule{\noalign{\hrule height\thinsize depth0pt}}%
\thicksize=1.5pt
\thinsize=0.6pt
\def\thickrule{\noalign{\hrule height\thicksize depth0pt}}%
\def\hrulefill{\leaders\hrule\hfill}%
\def\bigrulefill{\leaders\hrule height\thicksize depth0pt \hfill}%
\def\ctr#1{\hfil\ #1\hfil}%
\def\altctr#1{\hfil #1\hfil}%
\def\vctr#1{\hfil\vbox to0pt{\vss\hbox{#1}\vss}\hfil}%
%
%
\tablewidth=-\maxdimen%
\spreadwidth=-\maxdimen%
\def\tabskipglue{0pt plus 1fil minus 1fil}%
%
%
\centertablestrue%
\def\centeredtables{%
   \centertablestrue%
}%
\def\noncenteredtables{%
   \centertablesfalse%
}%
%
%
\parasize=4in%
\long\def\para#1{
   {%
      \vtop{%
         \hsize=\parasize%
         \baselineskip14pt%
         \lineskip1pt%
         \lineskiplimit1pt%
         \noindent #1%
         \vrule width0pt depth6pt%
      }%
   }%
}%
\gdef\ARGS{########}
\gdef\headerARGS{####}
\def\@mpersand{&}
{\catcode`\|=13
\gdef\letbarzero{\let|0}
\gdef\letbartab{\def|{&&}}%
\gdef\letvbbar{\let\vb|}%
}
{\catcode`\&=4
\def\ampskip{&\omit\hfil&}
\catcode`\&=13
\let&0
\xdef\letampskip{\def&{\ampskip}}%
\gdef\letnovbamp{\let\novb&\let\tab&}
}
\def\begintable{
   \begingroup%
   \catcode`\|=13\letbartab\letvbbar%
   \catcode`\&=13\letampskip\letnovbamp%
   \def\multispan##1{
      \omit \mscount##1%
      \multiply\mscount\tw@\advance\mscount\m@ne%
      \loop\ifnum\mscount>\@ne \sp@n\repeat%
   }
   \def\|{%
      &\omit\widevline&%
   }%
   \ruledtable
}
\long\def\ruledtable#1\endtable{%
%
%
%
   \offinterlineskip
   \tabskip 0pt
   \def\widevline{\vrule width\thicksize}
   \def\endrow{\@mpersand\omit\hfil\crnorm\@mpersand}%
   \def\crthick{\@mpersand\crnorm\thickrule\@mpersand}%
   \def\crthickneg##1{\@mpersand\crnorm\thickrule
          \noalign{{\skip0=##1\vskip-\skip0}}\@mpersand}%
   \def\crnorule{\@mpersand\crnorm\@mpersand}%
   \def\crnoruleneg##1{\@mpersand\crnorm
          \noalign{{\skip0=##1\vskip-\skip0}}\@mpersand}%
   \let\nr=\crnorule
   \def\endtable{\@mpersand\crnorm\thickrule}%
   \let\crnorm=\cr
%
%
   \edef\cr{\@mpersand\crnorm\tablerule\@mpersand}%
   \def\crneg##1{\@mpersand\crnorm\tablerule
          \noalign{{\skip0=##1\vskip-\skip0}}\@mpersand}%
   \let\ctneg=\crthickneg
   \let\nrneg=\crnoruleneg
   \the\tableLETtokens
%
%
   \tabletokens={&#1}
%
%
   \countROWS\tabletokens\into\nrows%
   \countCOLS\tabletokens\into\ncols%
%
%
   \advance\ncols by -1%
   \divide\ncols by 2%
   \advance\nrows by 1%
%
%
   \iftableinfo %
      \immediate\write16{[Nrows=\the\nrows, Ncols=\the\ncols]}%
   \fi%
%
%
   \ifcentertables
      \ifhmode \par\fi
      \hbox to \hsize{
      \hss
   \else %
      \hbox{%
   \fi
      \vbox{%
         \makePREAMBLE{\the\ncols}
         \edef\next{\preamble}
         \let\preamble=\next
         \makeTABLE{\preamble}{\tabletokens}
      }
      \ifcentertables \hss}\else }\fi
   \endgroup
   \tablewidth=-\maxdimen
   \spreadwidth=-\maxdimen
}
\def\makeTABLE#1#2{
   {
   \let\ifmath0
   \let\header0
   \let\multispan0
%
%
   \ncase=0%
   \ifdim\tablewidth>-\maxdimen \ncase=1\fi%
   \ifdim\spreadwidth>-\maxdimen \ncase=2\fi%
   \relax
%
   \ifcase\ncase %
      \widthspec={}%
   \or %
      \widthspec=\expandafter{\expandafter t\expandafter o%
                 \the\tablewidth}%
   \else %
      \widthspec=\expandafter{\expandafter s\expandafter p\expandafter r%
                 \expandafter e\expandafter a\expandafter d%
                 \the\spreadwidth}%
   \fi %
   \xdef\next{
      \halign\the\widthspec{%
      #1
      \noalign{\hrule height\thicksize depth0pt}
      \the#2\endtable
%
      }
   }
   }
   \next
}
\def\makePREAMBLE#1{
   \ncols=#1
   \begingroup
   \let\ARGS=0
   \edef\xtp{\widevline\ARGS\tabskip\tabskipglue%
   &\ctr{\ARGS}\tstrut}
   \advance\ncols by -1
   \loop
      \ifnum\ncols>0 %
      \advance\ncols by -1%
      \edef\xtp{\xtp&\vrule width\thinsize\ARGS&\ctr{\ARGS}}%
   \repeat
   \xdef\preamble{\xtp&\widevline\ARGS\tabskip0pt%
   \crnorm}
   \endgroup
}
\def\countROWS#1\into#2{
   \let\countREGISTER=#2%
   \countREGISTER=0%
   \expandafter\ROWcount\the#1\endcount%
}%
\def\ROWcount{%
   \afterassignment\subROWcount\let\next= %
}%
\def\subROWcount{%
   \ifx\next\endcount %
      \let\next=\relax%
   \else%
      \ncase=0%
      \ifx\next\cr %
         \global\advance\countREGISTER by 1%
         \ncase=0%
      \fi%
      \ifx\next\endrow %
         \global\advance\countREGISTER by 1%
         \ncase=0%
      \fi%
      \ifx\next\crthick %
         \global\advance\countREGISTER by 1%
         \ncase=0%
      \fi%
      \ifx\next\crnorule %
         \global\advance\countREGISTER by 1%
         \ncase=0%
      \fi%
      \ifx\next\crthickneg %
         \global\advance\countREGISTER by 1%
         \ncase=0%
      \fi%
      \ifx\next\crnoruleneg %
         \global\advance\countREGISTER by 1%
         \ncase=0%
      \fi%
      \ifx\next\crneg %
         \global\advance\countREGISTER by 1%
         \ncase=0%
      \fi%
      \ifx\next\header %
         \ncase=1%
      \fi%
      \relax%
      \ifcase\ncase %
         \let\next\ROWcount%
      \or %
         \let\next\argROWskip%
      \else %
      \fi%
   \fi%
   \next%
}
\def\counthdROWS#1\into#2{%
\dvr{10}%
   \let\countREGISTER=#2%
   \countREGISTER=0%
\dvr{11}%
\dvr{13}%
   \expandafter\hdROWcount\the#1\endcount%
\dvr{12}%
}%
\def\hdROWcount{%
   \afterassignment\subhdROWcount\let\next= %
}%
\def\subhdROWcount{%
   \ifx\next\endcount %
      \let\next=\relax%
   \else%
      \ncase=0%
      \ifx\next\cr %
         \global\advance\countREGISTER by 1%
         \ncase=0%
      \fi%
      \ifx\next\endrow %
         \global\advance\countREGISTER by 1%
         \ncase=0%
      \fi%
      \ifx\next\crthick %
         \global\advance\countREGISTER by 1%
         \ncase=0%
      \fi%
      \ifx\next\crnorule %
         \global\advance\countREGISTER by 1%
         \ncase=0%
      \fi%
      \ifx\next\header %
         \ncase=1%
      \fi%
\relax%
      \ifcase\ncase %
         \let\next\hdROWcount%
      \or%
         \let\next\arghdROWskip%
      \else %
      \fi%
   \fi%
   \next%
}%
{\catcode`\|=13\letbartab
\gdef\countCOLS#1\into#2{%
   \let\countREGISTER=#2%
   \global\countREGISTER=0%
   \global\multispancount=0%
   \global\firstrowtrue
   \expandafter\COLcount\the#1\endcount%
   \global\advance\countREGISTER by 3%
   \global\advance\countREGISTER by -\multispancount
}%
\gdef\COLcount{%
   \afterassignment\subCOLcount\let\next= %
}%
{\catcode`\&=13%
\gdef\subCOLcount{%
   \ifx\next\endcount %
      \let\next=\relax%
   \else%
      \ncase=0%
      \iffirstrow
         \ifx\next& %
            \global\advance\countREGISTER by 2%
            \ncase=0%
         \fi%
         \ifx\next\span %
            \global\advance\countREGISTER by 1%
            \ncase=0%
         \fi%
         \ifx\next| %
            \global\advance\countREGISTER by 2%
            \ncase=0%
         \fi
         \ifx\next\|
            \global\advance\countREGISTER by 2%
            \ncase=0%
         \fi
         \ifx\next\multispan
            \ncase=1%
            \global\advance\multispancount by 1%
         \fi
         \ifx\next\header
            \ncase=2%
         \fi
         \ifx\next\cr       \global\firstrowfalse \fi
         \ifx\next\endrow   \global\firstrowfalse \fi
         \ifx\next\crthick  \global\firstrowfalse \fi
         \ifx\next\crnorule \global\firstrowfalse \fi
         \ifx\next\crnoruleneg \global\firstrowfalse \fi
         \ifx\next\crthickneg  \global\firstrowfalse \fi
         \ifx\next\crneg       \global\firstrowfalse \fi
      \fi
\relax
      \ifcase\ncase %
         \let\next\COLcount%
      \or %
         \let\next\spancount%
      \or %
         \let\next\argCOLskip%
      \else %
      \fi %
   \fi%
   \next%
}%
\gdef\argROWskip#1{%
   \let\next\ROWcount \next%
}
\gdef\arghdROWskip#1{%
   \let\next\ROWcount \next%
}
\gdef\argCOLskip#1{%
   \let\next\COLcount \next%
}
}
}
\def\spancount#1{
   \nspan=#1\multiply\nspan by 2\advance\nspan by -1%
   \global\advance \countREGISTER by \nspan
   \let\next\COLcount \next}%
\def\dvr#1{\relax}%
\def\header#1{%
\dvr{1}{\let\cr=\@mpersand%
\hdtks={#1}%
\counthdROWS\hdtks\into\hdrows%
\advance\hdrows by 1%
\ifnum\hdrows=0 \hdrows=1 \fi%
\dvr{5}\makehdPREAMBLE{\the\hdrows}%
\dvr{6}\getHDdimen{#1}%
{\parindent=0pt\hsize=\hdsize{\let\ifmath0%
\xdef\next{\valign{\headerpreamble #1\crnorm}}}\dvr{7}\next\dvr{8}%
}%
}\dvr{2}}
\def\makehdPREAMBLE#1{
\dvr{3}%
\hdrows=#1
{
\let\headerARGS=0%
\let\cr=\crnorm%
\edef\xtp{\vfil\hfil\hbox{\headerARGS}\hfil\vfil}%
\advance\hdrows by -1
\loop
\ifnum\hdrows>0%
\advance\hdrows by -1%
\edef\xtp{\xtp&\vfil\hfil\hbox{\headerARGS}\hfil\vfil}%
\repeat%
\xdef\headerpreamble{\xtp\crcr}%
}
\dvr{4}}
\def\getHDdimen#1{%
\hdsize=0pt%
\getsize#1\cr\end\cr%
}
\def\getsize#1\cr{%
\endsizefalse\savetks={#1}%
\expandafter\lookend\the\savetks\cr%
\relax \ifendsize \let\next\relax \else%
\setbox\hdbox=\hbox{#1}\newhdsize=1.0\wd\hdbox%
\ifdim\newhdsize>\hdsize \hdsize=\newhdsize \fi%
\let\next\getsize \fi%
\next%
}%
\def\lookend{\afterassignment\sublookend\let\looknext= }%
\def\sublookend{\relax%
\ifx\looknext\cr %
\let\looknext\relax \else %
   \relax
   \ifx\looknext\end \global\endsizetrue \fi%
   \let\looknext=\lookend%
    \fi \looknext%
}%
%
%
\def\tablelet#1{%
   \tableLETtokens=\expandafter{\the\tableLETtokens #1}%
}%
\catcode`\@=12

\vspace{-10mm}
\section{Introduction}

The theoretical tools available at present are not sufficient
to understand hadronic collisions at high energies from QCD alone.  
Phenomenological models of so-called \textit{soft} multiparticle
production are typically applied in addition to perturbative QCD. The Dual Parton
Model (DPM) \cite{Capella94a} is such a  {phenomenological}
model. It is quite successful in its understanding of many details in the  multiparticle production. Its fundamental ideas are presently 
the basis of many of the Monte Carlo implementations of soft interactions.

The properties of  our DPM implementation 
\textsc{DPMJET}--III~ are described in
\cite{Engel95a,Engel95d,Roesler20001,Roesler20002,posterpap04}. For a more detailed description of
\textsc{DPMJET}--III we refer to these papers and the literature quoted
therein.

A feature of hadron production in  nuc\-lear
collisions discussed in the last 10 years is the large stopping of the
participating nucleons 
in hadron--nucleus and nucleus--nucleus collisions. Experimental data clearly 
demonstrating the sizable stopping of the  participating nuc\-leons in
hadron--nucleus and nucleus--nucleus interactions for fixed-target
experiments have been presented in \cite{NA35FIN,Alber98,Aitala2000} and
\cite{NA49pp158}. 
 
Multistring fragmentation models like the Dual Parton Mo\-del (DPM)
or similar models contain some stopping, but in their original form they did not account
for the enhanced stopping found in nuclear collisions.
Therefore, in order to incorporate the effect into multistring
fragmentation mo\-dels
 new diquark breaking DPM--diagrams acting in
 hadron--nucleus and nucleus--nucleus collisions were 
 proposed by \cite{Kharzeev96} and 
 \cite{Capella96} and investigated in detail in
 \cite{Capella99a} and \cite{Capella99}. 
 Similar ideas were discussed by \cite{Vance99} and \cite{Casado99}.
The  Monte Carlo implementation into DPMJET--II.5 of the new
diquark breaking diagrams of \cite{Kharzeev96} and 
  \cite{Capella96} was discussed in \cite{Ranft20001}. 
The implementation into  DPMJET--III~\cite{posterpap04} 
of these diagrams differs somewhat from that of
\cite{Ranft20001} and was described in \cite{Ranft20004,Ranft20003}.

A second mechanism, 
which can contribute  
to baryon stopping in nuclear collisions 
occurs during the fusion of 
chains \cite{Braun97,Braun2000a}. This mechanism was introduced into
DPMJET--III in order to accommodate the RHIC data
\cite{Ranft2003,Bopp2004}. We will  use chain fusion here exactly as
described in these papers.

Particle production ratios in p--p and d--Au collisions measured at RHIC
\cite{BackPHOBOS03,SteinbergPHOBOS04,STAR0412,BRAHMS0409} allow us a more
precise determination of the parameters of anomalous baryon stopping. 
Needed stopping contribution are described in section II. Sections III and
IV considere the available data on stopping for proton-proton,
deuteron-gold reactions. In section V we considere hyperon productions
ratioes in pion-proton collisions. There we need another new mechanism to
describe the data.
\section{Implementation of new diagrams for an improved
description of baryon stopping}

As justified by  Rossi and Veneziano~\cite{Rossi97} 
in the framework of a $1/N$ expansion a 
baryon can
be pictured as made out of three quarks bound together by three
strings which join in a so-called string junction point.
In diagrams one can characterize the baryons \vspace*{-2mm}
\begin{description}
\item(i) by the three quarks and the string junction or\vspace*{-2mm}
\item(ii) by a quark and a diquark. \vspace*{-2mm}

\end{description} 
\noindent
In the second  case the string junction always goes with the diquark.

Figs.~1-4 are quark line diagrams describing the production chains of
particles.
In  these diagrams   quarks and diquarks are usually plotted as 
single resp.\ paired solid lines. 
If diquarks break,  string junctions are added as dashed lines.

\begin{figure}[thb] \centering
\begin{turn}{-90}
\vspace*{-10mm}
\includegraphics[height=6cm,width=4cm]{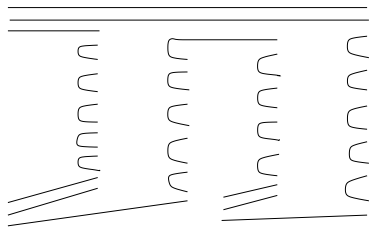}

\end{turn}
 \vspace*{2mm}
 \caption{The diquark--conserving diagram for a nucleon--nucleus
 collision with two participants of the target nucleus.
 \protect\label{diqc}
 }

 \end{figure}

\begin{figure}[thb] \centering 
\vspace*{-10mm}
\begin{turn}{-90}
\includegraphics[height=6cm,width=4cm]{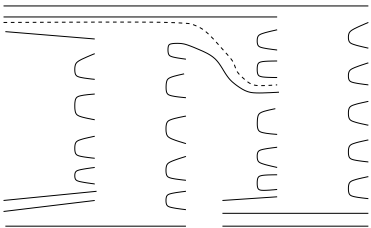}
\end{turn}
 \vspace*{2mm}
 \caption{The Glauber sea quark mechanism of baryon stopping GSQBS for a
 nucleon--nucleus collision with two 
 participants of the target nucleus. This
 is the second C--K diquark--breaking mechanism \protect\cite{Capella96}.
 \protect\label{diqsea}
 }
 \end{figure}

In the fragmentation of diquarks
there are  two possibilities for the first fragmentation step. 
 Either one gets a baryon, which contains the diquark (with 
 the string junction) in the first step 
or one gets a meson containing only one of the two quarks. 
In this case the
string junction migrates to the next step and the
baryon is eventually produced in one of  
the following fragmentation steps. 

This mechanism is well-known, it is presented in the review on the
Dual Parton Model
\cite{Capella94a} and it was investigated 
for instance in \cite{SUKHA,KOPEL}.
 This mechanism was implemented
a long time ago (1980) in the 
BAMJET--fragmentation code \cite{BAMJET1,BAMJET}
used in the early versions of DPMJET. 
This mechanism is also implemented under the name {\it popcorn}
fragmentation in the Lund chain fragmentation model JETSET
\cite{JETSET,AND85} which is presently used in DPMJET.

 What happens in the model with the popcorn mechanism compared to the
model without can be most easily seen by looking at the proton
rapidity distribution in p--p collisions.
The two maxima in the target and projectile
fragmentation region of the proton rapidity distribution shift by
about half a unit to the center,  the peaks become
wider and  correspondingly the dip in the center is reduced.
At the same time the Feynman $x$ distributions of mesons get
a component at larger Feynman $x$. The effects in hadron--nucleus and
nucleus--nucleus collisions are quite similar.
However, the popcorn mechanism  alone cannot explain the baryon
stopping observed experimentally in hadron--nucleus and nucleus--nucleus
collisions \cite{NA35FIN,Alber98}.

\subsection{Nuclear diquark breaking}


Most interesting for DPMJET is the so-called {\it second C--K}
mechanism~\cite{Capella96,Capella99a,Capella99,Capella99b}.
In Fig.~\ref{diqc} we plot first the diquark--conserving diagram for a
nucleon--nucleus collisions with two participants of the target
nucleus. This is the traditional way for such a collision in the DPM.
In Fig.~\ref{diqsea} we plot the second C--K
diquark--breaking diagram for the same collision. Now the second 
valence quark from the broken diquark is replaced by a Glauber 
sea quark from the nucleon projectile. Therefore, we will
call the mechanism the Glauber 
sea quark mechanism of baryon stopping (GSQBS). The probability of such a
diquark splitting rises if the considered nucleon is involved in more
than two interactions.

A new type of string combination appears in the second interaction,
which happened between the second valence quark (the top one)  
and the junction line. 
A central assumption is that a sea quark of the initial
scattering determines the position of the junction line. 
One of the valence quarks  which its initial momentum connects to the junction 
line by an upward string. In this way the string pulling the
vortex line down is compensated by a string pulling it up.

The GSQBS picture is chosen for simplicity. 
Formally at the very top three strings and a vortex line 
are exchanged. In the topological view it is a cut 
t-channel-quarkless-baryonium-exchange. 
Usually baryonium exchanges have a rather low intercept 
and the idea is that in special  situations a small 
quarkless component with a high intercept appears. 
Taking the position from the sea quark a flat distribution
of the vortex line with a intercept of one is implied, 
with an added bias to larger values as the sea quark 
actually had to sit on the forward end of string of a minimum size. 
In the factorizing version of the model\cite{Boppshabelski} 
this bias does not exist. 
The intercept is then estimated to be slightly less than one.

The GSQBS has been
implemented into DPMJET--II.5 and DPMJET--III. 
With this mechanism one is able
to fill the dip in the baryon rapidity distributions at central rapidity
in agreement to the experimental data. As discussed already in detail in
\cite{Capella99,Capella99b} 
this mechanism also contributes to increase the Hyperon
production in nucleon--nucleus and nucleus--nucleus collisions.

\subsection{Hadronic diquark breaking at high energy}

At high energies multiple collisions appear even in hadron--hadron
scattering due to the unitarization procedure. This has the  
consequence that new diagrams like the GSQBS diagrams become necessary.
In some way or other such diagramms have to be implemented  in any model 
which includes both elastic and inelastic processes. 

We call the sea quarks at the ends of the additional chains in 
this case {\it unitary sea quarks}. They are relevant for high energies.
The Glauber sea quarks are needed in nuclear collisions already at rather
low energies, for instance at the energies of heavy ion collisions at
the CERN--SPS. In contrast to this, unitary sea quarks appear in
significant numbers only at rather high energies, for
instance at the energies of RHIC, the CERN--SPS collider or the Tevatron
collider.

\begin{figure}[thb] \centering
\vspace*{-10mm}
\begin{turn}{-90}
\includegraphics[height=7cm,width=4cm]{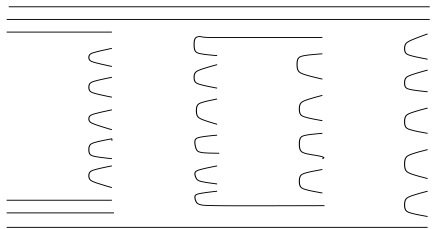}
\end{turn}\hspace*{30mm}
 \vspace*{2mm}
 \caption{Standard DPM diagram for a nucleon--nucleon interaction with
 one additional soft secondary interaction 
 induced by the unitarization procedure.
 \protect\label{normal2}
 }
 \end{figure}

\begin{figure}[thb] \centering
\vspace*{-10mm}
\begin{turn}{-90}
\includegraphics[height=7cm,width=4cm]{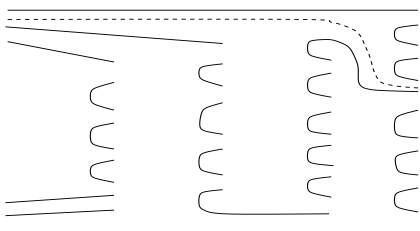}
\end{turn}\hspace*{30mm}
 \vspace*{2mm}
 \caption{New DPM diagram for a nucleon--nucleon interaction with one
 additional soft secondary interaction 
 induced by the unitarization procedure. 
 We call this 
 unitary sea quark mechanism for 
 baryon stopping USQBS \cite{Ranft20001}.
 \protect\label{new2}
 }
 \end{figure}

With the unitary sea
quarks at the ends of the chains from the secondary collisions one obtains
a new mechanism for baryon stopping which will become effective at very
high energies. It is illustrated in Figs.~\ref{normal2} and \ref{new2}.

The standard DPM diagram is plotted in Fig.~\ref{normal2} for a
nucleon--nucleon interaction with two soft  interactions induced
by the unitarization procedure. 
As in Fig.~\ref{diqc}, there is one valence--valence 
and one sea--sea interaction, each
represented by a pair of chains. In analogy to Fig.~\ref{diqsea},
a new diagram \cite{Ranft20001} 
for baryon stopping is constructed in Fig.~\ref{new2}.
The diquark is split and a unitary sea quark is used 
 to have the  baryon only in the second or later fragmentation steps 
 in one of the chains. We call this the
 unitary sea quark mechanism for baryon stopping USQBS. 
 The implementation of the new diquark breaking diagrams in DPMJET--III
is discussed in detail in \cite{Ranft20004,Ranft20003} . The relative
probability of this process introduces one parameter, which will be
given later. 

Again the probability for such a diquark splitting rises if there are more than 2
 interactions of the hadrons involved. Obviously, 
 this mechanism leads to  Feynman $x$ distributions of baryons in
 p--p collisions becoming  softer and Feynman $x$ distributions of
 mesons becoming  harder than without the USQBS mechanism.

In contrast to the  GSQBS mechanism which leads already
to observable changes in nuclear collisions 
at the energy of the CERN--SPS, there were no data available before
RHIC  to prove that this USQBS mechanism is a needed
extension of the DPM. The situation concerning baryon stopping in
proton--proton or antiproton--proton collisions at lower energies
was inconclusive \cite{auren77}. An idea to observe baryon stopping in
diffractive events \cite{bopp2000}, which always contain two
interactions, was unfortunately not  pursued in experimental analysis.

\subsection{New parameters connected with the diquark breaking
diagrams
}

For each of the new diquark breaking diagrams described in this section
a new parameter has to be introduced. These parameters give the
probability for the diquark breaking mechanisms to occur, 
given a suitable sea
quark is available and given that the diquark breaking mechanism is
kinematically allowed. For an original diquark--quark chain of small
invariant mass, which originally 
just fragments into two hadrons, the diquark
breaking is often not allowed at small energies.

The values of the new parameters are determined by comparing
DPMJET--III with experimental data on  antiparticle to particle
production ratios and on
net--baryon distributions.

\section{  Antiparticle to particle ratios in proton--proton collisions
 }
 
With antibaryon to baryon ratios measured at RHIC one is now 
for the first time in a position to determine the USQBS parameter with good accuracy. 
In p--p collisions, the contribution of chain fusion is not very large.

 \begin{figure}[thb] \centering
\vspace{-2mm}
\includegraphics[height=7cm,width=8cm]{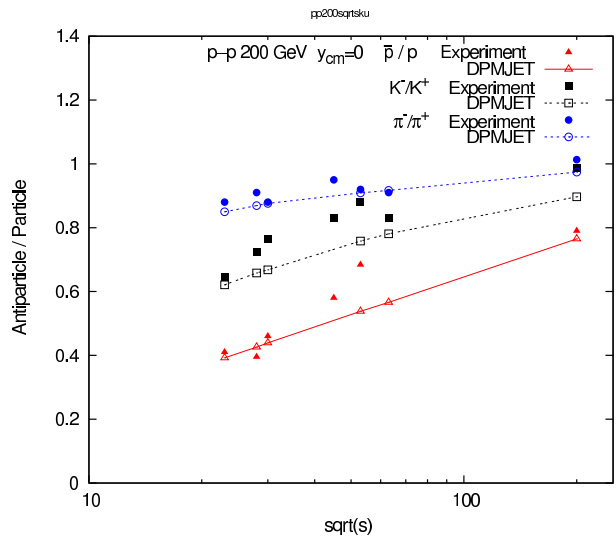}
\vspace{-2mm}

\caption{
$\sqrt s$ dependence of antiparticle to particle ratios in p--p
collisions at $y_{cm}$ = 0. Experimental data are from the BRAHMS
Collaboration at RHIC \cite{BRAHMS0409}, the ISR \cite{BBC75} and the
NA27 Collaboration \cite{Aguilar91a}. 
We compare \textsc{Dpmjet}-III results with the experimental data.
}

\vspace{-2mm}
\label{fig:pp200sqrtsku}
\end{figure}

\begin{figure}[thb]
\begin{center}
\vspace{-2mm}
\vspace*{-4mm}
\hspace*{-15mm}\includegraphics[height=6cm,width=7cm]{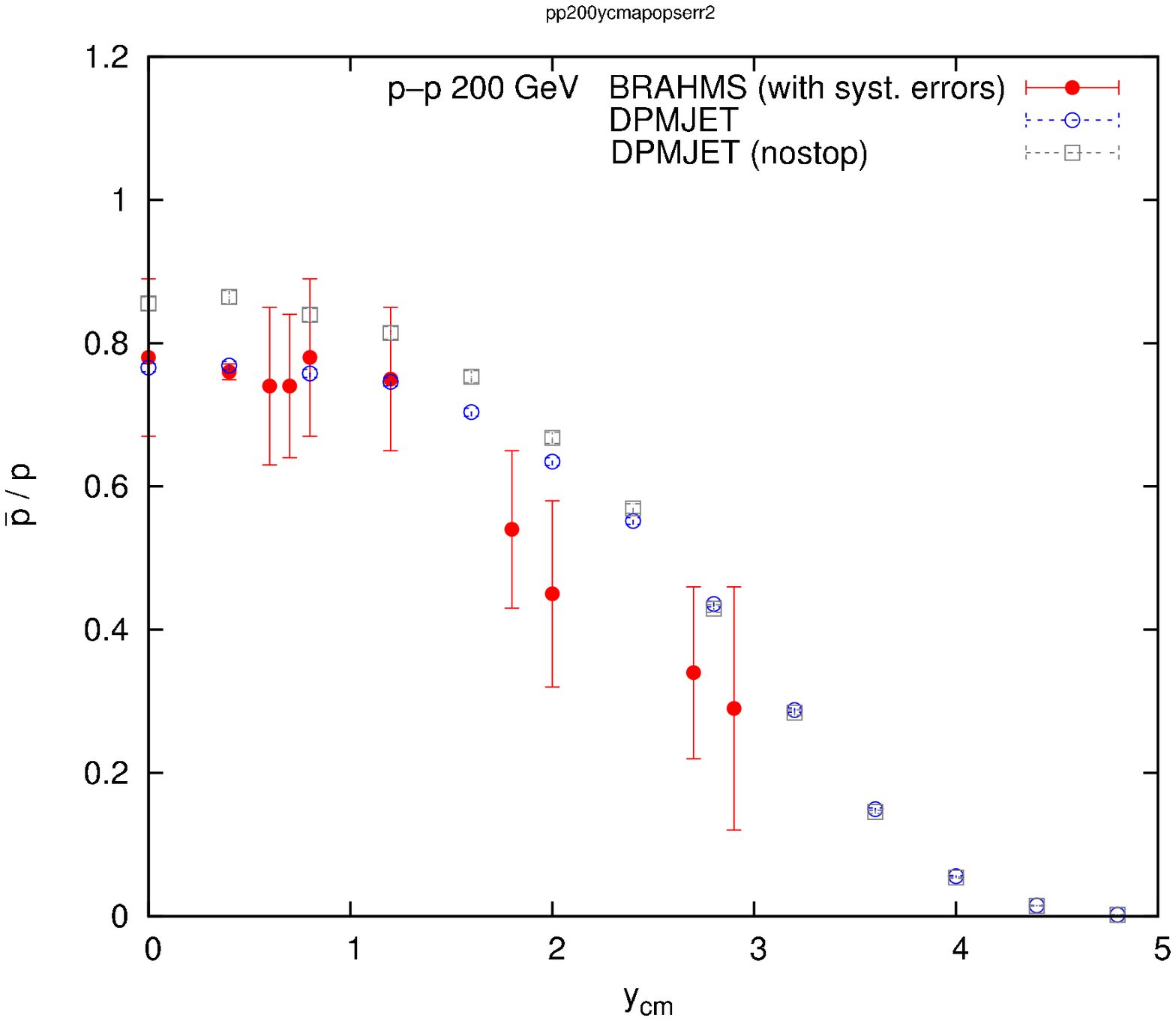}\hspace*{20mm}\\
\hspace*{-15mm}\includegraphics[height=6cm,width=7cm]{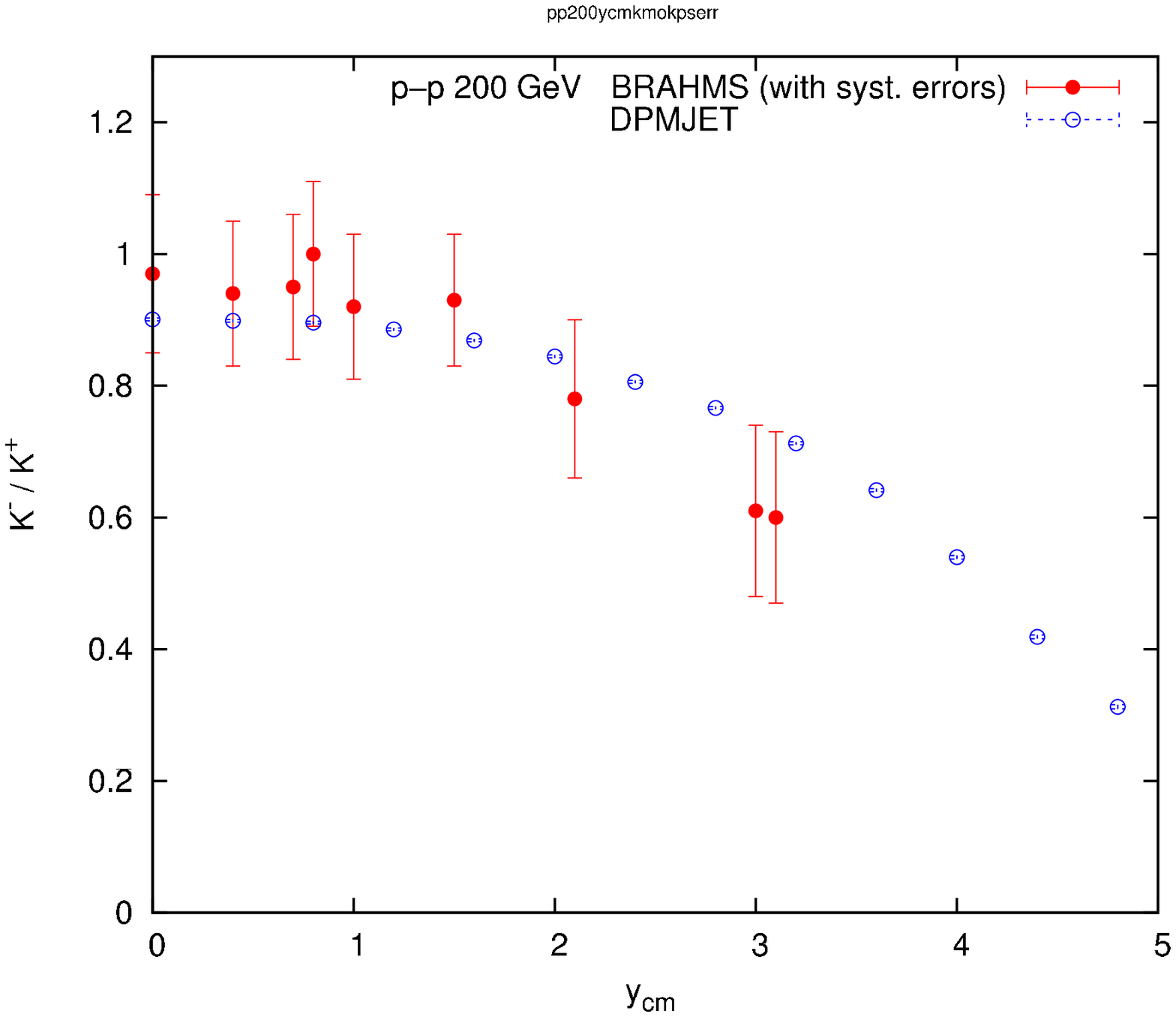}\hspace*{20mm}\\
\hspace*{-15mm}\includegraphics[height=6cm,width=7cm]{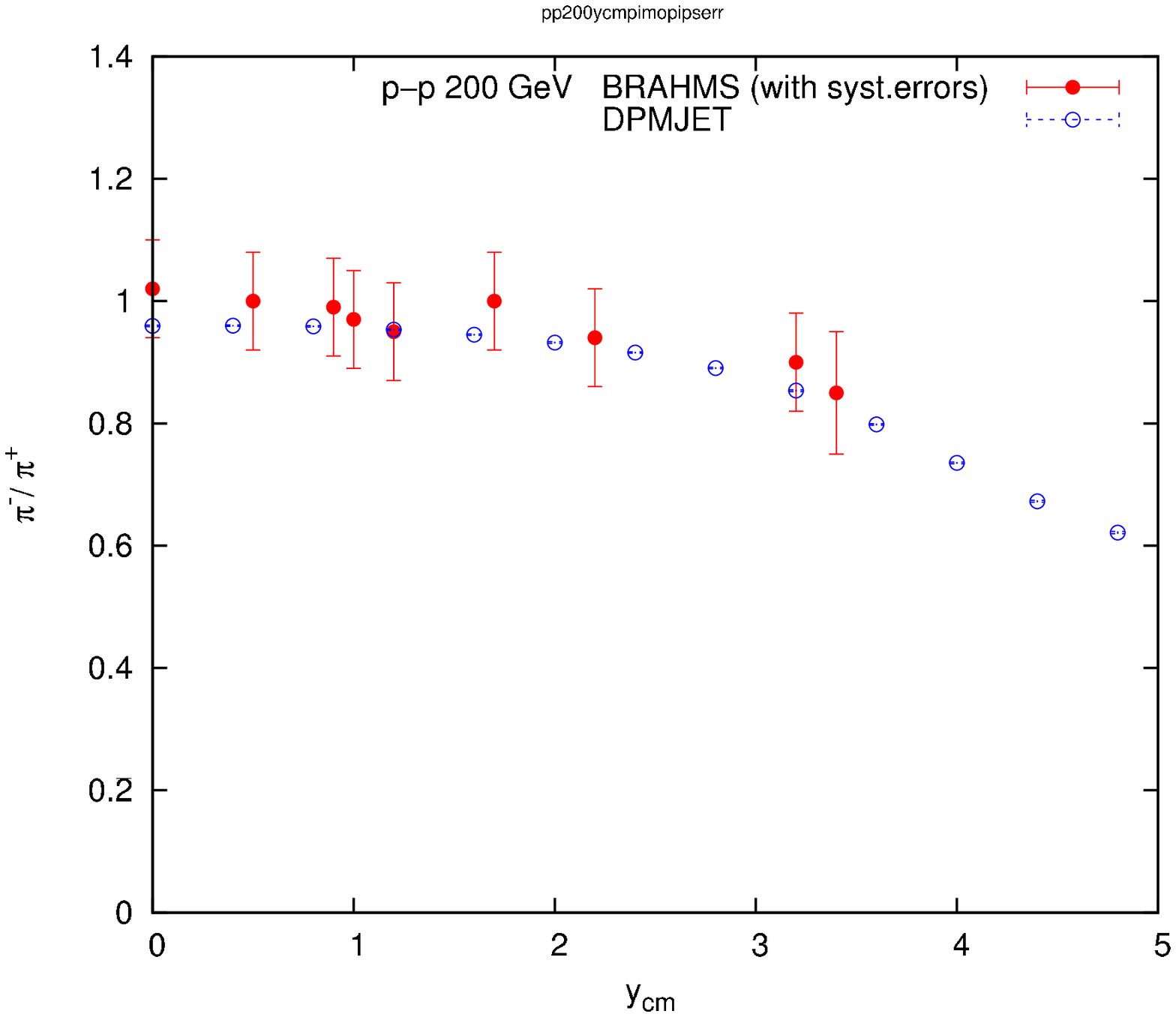}\hspace*{20mm}\\
\end{center}
\vspace{-3mm}
\caption{
Antiparticle to particle ratios ($\bar p/p$,  $K^-$/$K^+$ and $\pi ^-$/$\pi ^+$) in p--p collisions at 200 GeV as function of
the cms--rapidity $y_{cm}$. We compare data from the BRAHMS Collaboration at
RHIC \cite{BRAHMS0409} (including the systematic errors as given in the
BRAHMS paper) with the results obtained from 
 \textsc{Dpmjet}-III. We plot the $\bar p$/p ratios for the full DPMJET
 model as well as for the model without the USQBS
 diagrams("nostop").
 }

\vspace{-6mm}
\label{fig:pp200ycmapopserr}\label{fig:pp200ycmkmokpserr}\label{fig:pp200ycmpimopipserr}
\end{figure} 

Fig.~\ref{fig:pp200sqrtsku} compares measured $\pi ^-$/$\pi ^+$,
$K^-$/$K^+$ and $\bar p$/$p$  ratios at $y_{cm}$ = 0 as function
of the energy with the DPMJET--III results, of course the USQBS
mechanism influences essentially only the antiproton to proton ratio but
also the  comparison of the antimeson to meson ratios is of interest.

We get  agreement to the data for  $\pi ^-$/$\pi ^+$ and 
$K^-$/$K^+$. To obtain  agreement with the $\bar p$/$p$ ratio a USQBS parameter of .07 
had to be chosen to get a $0.1$ reduction at the highest energy. It means that in 7 \% of the possible
cases one transforms the diquark conserving diagram as given 
in Fig.~\ref{normal2} into the diquark breaking diagram given in 
Fig.~\ref{new2}. It corresponds to the parameter $\epsilon$ of \cite{Arakelian:2002iw,Boppshabelski} whose stability 
against variation in the model was investigated in the cited papers.


%
%
%
%
%
%
%
%
%
The dependence  of the three antiparticle to particle production ratios
on the cms rapidity of the produced particles is plotted in 
Fig.~\ref{fig:pp200ycmapopserr}.
Here we have kept the USQBS
parameters as determined above in this comparison of DPMJET--III
with the BRAHMS data from RHIC \cite{BRAHMS0409}.  BRAHMS data 
are plotted with statistical and systematic errors 
as given in \cite{BRAHMS0409}.
In Fig.~\ref{fig:pp200ycmapopserr}.a we compare the BRAHMS data for $\bar
p/p$ ratios with the full DPMJET model as well as with the model without
the USQBS diagrams. It shows that the 
USQBS diagrams are needed to get a better agreement with the 
BRAHMS data at central rapidity.
In all cases DPMJET--III describes the experimental data practically
within the errors.

\begin{figure}[h]
\begin{center}
\includegraphics[height=8cm,width=8cm]{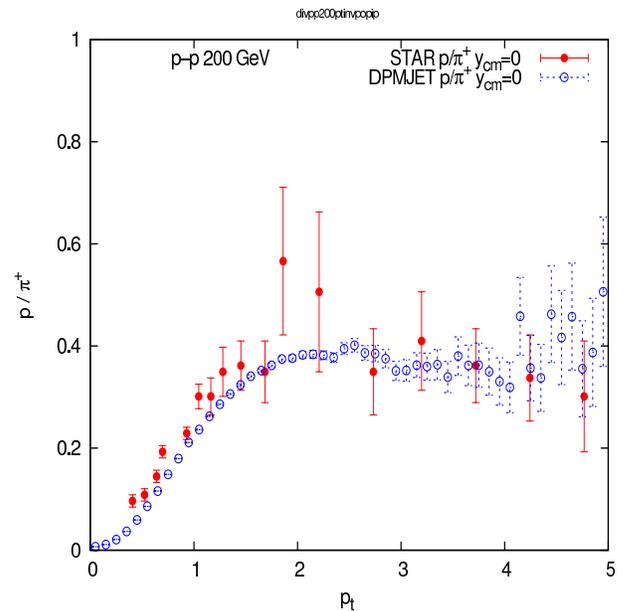}
\end{center}
\vspace{-7mm}

\caption{ The p/${\pi}^+$ particle production ratio as function of the
transverse momentum in $\sqrt s$ = 200 GeV proton--proton collisions. We
compare the ratio according to DPMJET-III with data from the
STAR--Collaboration \cite{STARptopip}.
 }

\vspace{-1mm}
\label{fig:divpp200ptinvpopip}
\end{figure}

In Fig.\ref{fig:divpp200ptinvpopip} we plot the p/${\pi}^+$ particle
production ratio as function of the
transverse momentum in $\sqrt s$ = 200 GeV proton--proton collisions.
The particle production ratio according to DPMJET-III is compared with
data from the
STAR--Collaboration \cite{STARptopip}. We find an excellent agreement.

 \begin{figure}[h]
\begin{center}
\includegraphics[height=7cm,width=8cm]{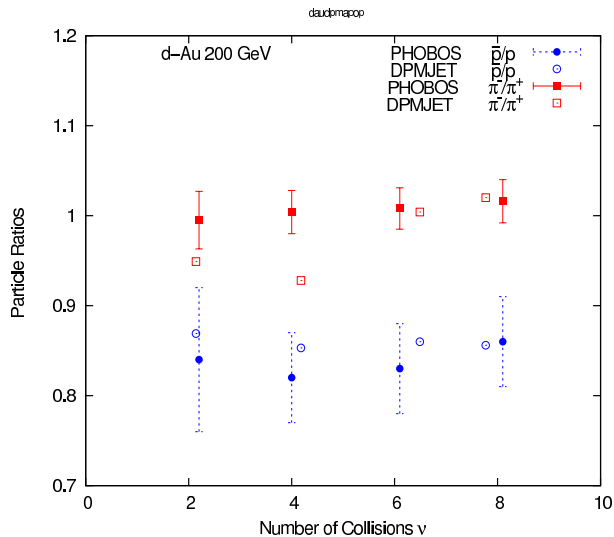}
\end{center}

\caption{
Centrality dependence of charged antiparticle to particle ratios
near mid--rapidity as measured by the PHOBOS--Collaboration
\cite{SteinbergPHOBOS04} in d--Au collisions at $\sqrt s$ = 200 GeV is
compared to the {\sc DPMJET}-III model.
 }

\label{fig:daudpmapop}
\end{figure}

\section{  Antiparticle to particle ratios in $\bf d-Au$ collisions
 }
In d--Au collisions we have, in addition to the baryon stopping
mechanisms acting in p--p collisions, also the GSQBS diagrams and the
contribution from chain fusion to baryon stopping.

 The centrality dependence of charged antiparticle to particle ratios
 near mid--rapidity was measured by the PHOBOS--Collaboration
 \cite{BackPHOBOS03} in d--Au collisions at $\sqrt s$ = 200 GeV as function of the centrality. 
The $\pi ^-$/$\pi ^+$ and $\bar p$/$p$ ratios are compared in
 Fig.~\ref{fig:daudpmapop}  with the {\sc Dpmjet}-III results.
 In this comparison
 the $\pi ^-$/$\pi ^+$ and  $\bar p$/$p$ ratios at all 
 centralities agree within the experimental
 errors, the statistical errors of the Monte Carlo calculations are
 below $\pm 0.02$.

 Regarding the exact position of the points, the
 measurements and also our {\sc DPMJET}-III calculations were done for
 four different centrality bins: 0-10\%, 10--30\%, 30--60\% and
 60--100\% but the resulting antiparticle to particle ratios are plotted
 as function of $\nu$, the average number of collisions per deuterium
 participants. PHOBOS finds for the four centralities given above $\nu$
 = 8.1, 6.1, 4.0 and 2.2, in the {\sc Dpmjet}-III calculations we find
 $\nu$ = 7.77, 6.49, 4.18 and 2.14.

 \begin{figure}[h]
\begin{center}
\hspace*{-15mm}\includegraphics[height=7cm,width=8cm]{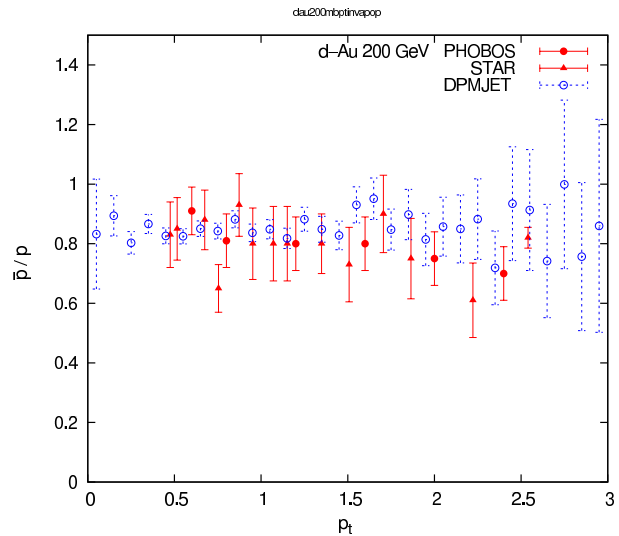}
\hspace*{20mm}\\
\end{center}

\label{fig:dau200mbptinvapop}
\begin{center}
\hspace*{-15mm}\includegraphics[height=7cm,width=8cm]{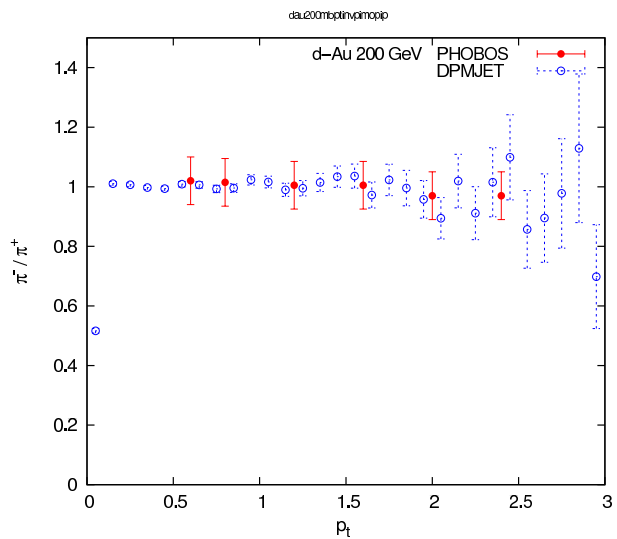}\hspace*{20mm}\\
\end{center}

\caption{
Transverse momentum dependence of the $\bar p$ to $p$ resp. $\pi ^-$ to $\pi ^+$ ratio in d--Au
collisions at $\sqrt s$ = 200GeV. We compare the data from the PHOBOS and
the STAR
Collaboration at RHIC  
 \cite{SteinbergPHOBOS04,AdamsSTAR06} 
  to the results from the {\sc Dpmjet}-III model.
 }

\vspace{-1mm}
\label{fig:dau200mbptinvpimopip}
\end{figure} 

In Fig.
\ref{fig:dau200mbptinvpimopip} 
we plot the $\bar p$/$p$ and
$\pi ^-$/$\pi ^+$ ratios in d--Au collisions at $\sqrt s$ = 200 GeV as
function of the transverse momentum. Compared are experimental data of
the PHOBOS and STAR Collaboration  \cite{SteinbergPHOBOS04,AdamsSTAR06} with the 
results from  {\sc Dpmjet}-III , we find a reasonable agreement.

 \begin{figure}[h]
\begin{center}
\includegraphics[height=7cm,width=8cm]{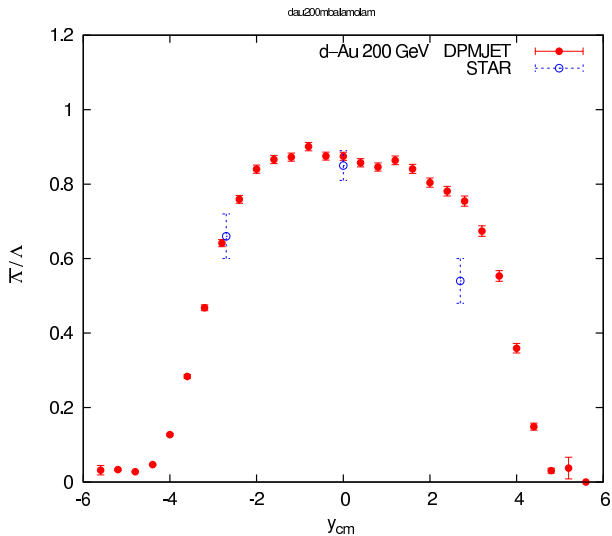}
\end{center}

\caption{
Anti--$\Lambda$ to $\Lambda$ ratio as function of the cms--rapidity in minimum
bias d--Au collisions at $\sqrt s$ = 200 GeV.
We compare the experimental data from the STAR Collaboration at RHIC
\cite{STAR0412}
  to the results from the {\sc Dpmjet}-III model.
 }

\label{fig:dau200mbalamolam}
\end{figure}

\begin{figure}[h]
\begin{center}
\includegraphics[height=7cm,width=8cm]{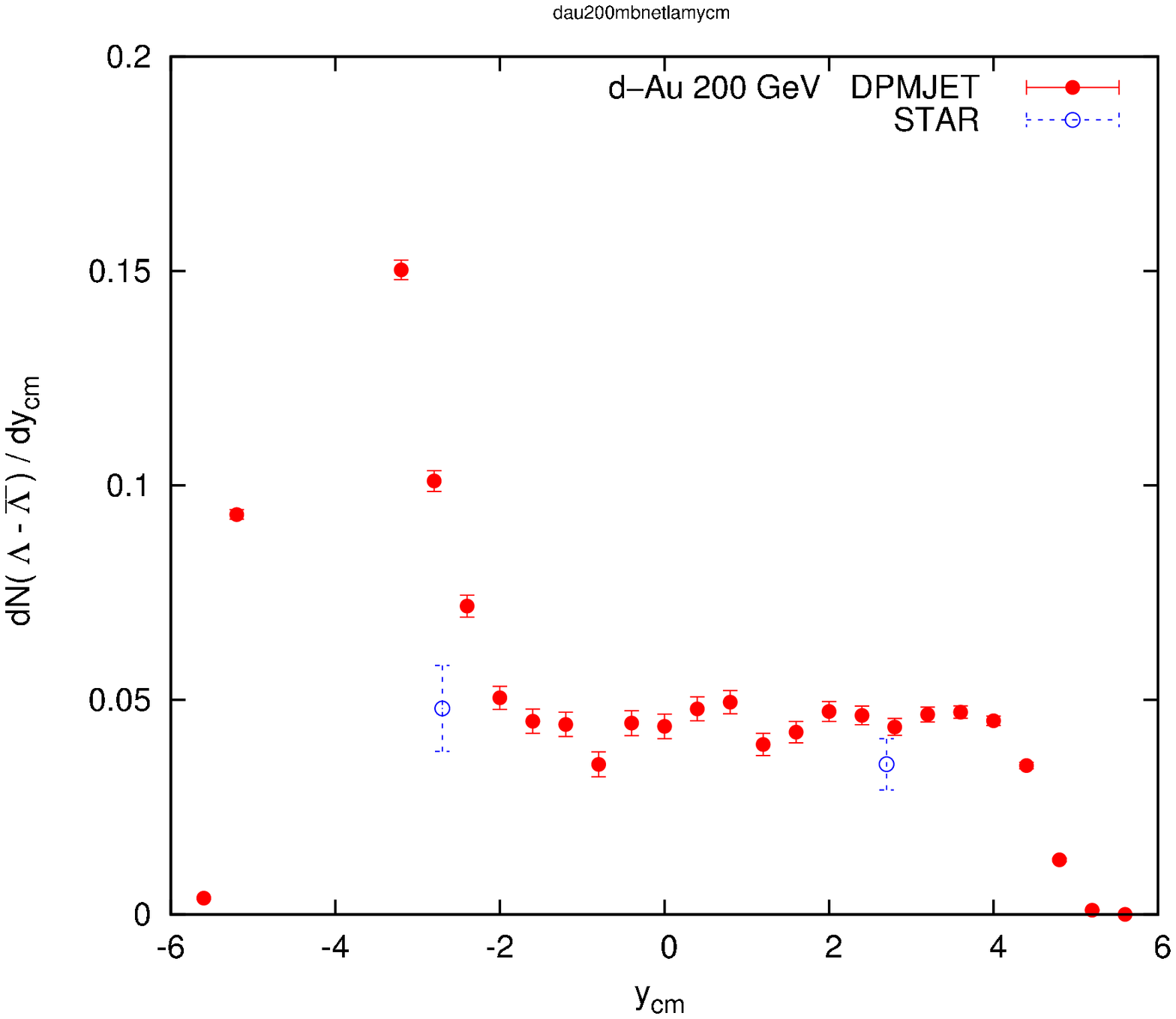}
\end{center}

\caption{
NET--$\Lambda$ production as function of the cms--rapidity in minimum
bias d--Au collisions at $\sqrt s$ = 200 GeV.
We compare the experimental data from he STAR Collaboration at RHIC
\cite{STAR0412}
  to the results from the {\sc Dpmjet}-III model.
 }

\vspace{-1mm}
\label{fig:dau200mbnetlamycm}
\end{figure} 

In Figs.
\ref{fig:dau200mbalamolam} and
\ref{fig:dau200mbnetlamycm} we plot $\bar \Lambda$/$\Lambda$ ratios and
net--$\Lambda$ production ($\Lambda$ - $\bar \Lambda$) in d--Au
collisions at $\sqrt s$ = 200 GeV as function of the cms rapidity.
Compared are the experimental data of the STAR Collaboration
\cite{STAR0412}
with the results of  
   the {\sc Dpmjet}-III model. 
The agreement for the ratios $\bar
\Lambda$/$\Lambda$ is satisfactory given the different binning 
, significant might be some disagreement for net
   $\Lambda$ production in the Au fragmentation region. For all {\sc
   Dpmjet}-III calculations in Figs.
\ref{fig:daudpmapop} to \ref{fig:dau200mbnetlamycm} we use {\sc
Dpmjet}-III with chain fusion as described in 
\cite{Ranft2003,Bopp2004}. 

The USQBS parameter is used exactly as
described in the previous section. 
The GSQBS parameter is fitted.   
It turned out to be rather small at this energy and was set 
to zero in the shown fits.

The vanishing of the GSQBS contribution
should not be taken too seriously. In nuclear
collision there are two baryon stopping mechanisms.
Besides  GSQBS baryon stopping is obtained as a side effect      
from chain fusion (e.g. a $qq-q$ chain and a $\bar q - q$ chain
can fuse to a $q-qq$) \cite{Ranft2003,Bopp2004}.     
Both the intricate string structure of GSQBS and the
forward-constituents-backward-constituents structure of fusion
strings are somewhat ad hoc and an redistribution between both
contributions is clearly acceptable.
More important is the fact
that the observed nuclear baryon stopping is within the 
expected range (see also \cite{Capella:2002sx}).

%
%


%

\section{ Antihyperon to Hyperon ratios in hadron--hadron collisions }

\begin{figure}[h]
\begin{center}
\includegraphics[height=3.0cm,width=8cm]{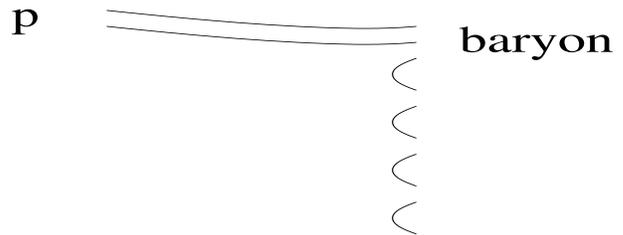}
\end{center}

\caption{ Fragmentation into a baryon, the most important term 
in the fragmentation of a diquark.
 }

\vspace{-1mm}
\label{fig:baryon-leading3}
\end{figure}

\begin{figure}[h]
\begin{center}
\includegraphics[height=3.5cm,width=9cm]{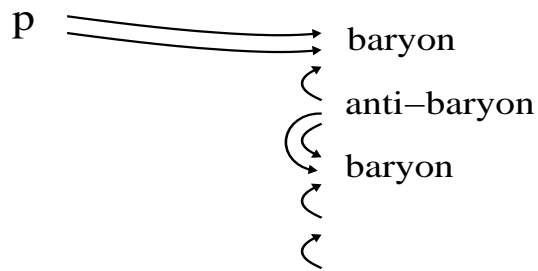}
\end{center}

\caption{Production of a antibaryon--baryon pair in the standard
fragmentation of a diquark. 
 }

\vspace{-1mm}
\label{fig:rank_3s_baryon3}
\end{figure}

\begin{figure}[h]
\begin{center}
\includegraphics[height=3.5cm,width=8cm]{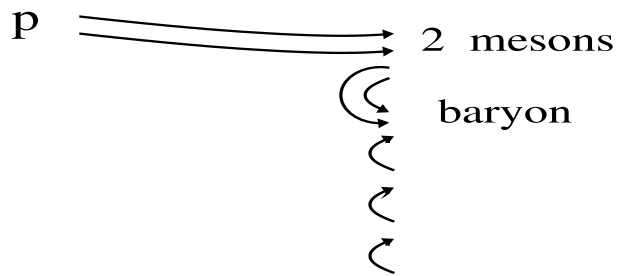}
\end{center}

\caption{Non--standard fragmentation of a diquark into a leading
meson--pair and a baryon.
 }

\vspace{-1mm}
\label{fig:rank_2s_baryon3}
\end{figure} 

Asymmetries of strange baryon production in 500 GeV $\pi^-$--p
collisions were measured by the E791 Collaboration at Fermilab
\cite{Aitala2000}.
The asymmetry is defined as follows
\begin{equation}
  A(\frac{B}{\bar B}) = \frac{N_B - N_{\bar B}}{N_B + N_{\bar B}}
\end{equation}
in each $x_F$ bin. These asymmeties are obviously closely related to
the particle production ratios $\frac{\bar B}{B}$.

All measured asymmeties \cite{Aitala2000} $A({\Lambda}/{\bar
\Lambda})$, $A({\Xi^-}/{{\bar \Xi}^+})$ and $A({\Omega^-}/{{\bar
\Omega}^+})$ are positive, correspondingly in the given $x_F$ range
the ratios ${\bar B}/{B}$ are smaller than 1, there are more
Hyperons produced than anti--Hyperons.

It was already noted by the E791 Collaboration in \cite{Aitala2000}
that the PYTHIA/JETSET code
\cite{JETSET,AND85} gives $A({\Omega^-}/{{\bar \Omega}^+})$ and  
$A({\Xi^-}/{{\bar \Xi}^+})$ asymmeties, which are negative or zero
in part of the kinematic range, correspondingly in part of the kinematic
range the PYTHIA/JETSET code predicts more double or triple strange
anti--Hyperons than Hyperons. This observation was also discussed in
detail by Liu et al. \cite{Liu2003}.

DPMJET--III uses PYTHIA/JETSET for the fragmentation of the hadronic
strings, the building blocks of the model. Therefore, we are not
surprised that also the original DPMJET-III gives
$A({\Omega^-}/{{\bar \Omega}^+})$ and
     $A({\Xi^-}/{{\bar \Xi}^+})$ asymmeties, which are negative
    or zero
  in part of the kinematic range in contrast to the experimental data.

  To find the reason for this wrong behaviour of chain
  fragmentation models like DPMJET--III or  PYTHIA we have to consider
  the mechanism for baryon (and especially double and triple strange
  hyperon) production in chain decay models.

In Fig.\ref{fig:baryon-leading3} we plot the standard diagram of chain
decay models for the
fragmentation of a diquark into a leading baryon. In collisions of
non--strange hadrons the chain--end diquarks do not contain strange
quarks, therefore only hyperons with one strange quark can be 
produced this way. These are  the $\Lambda$ hyperons in the E791
experiment. 

Next, in Fig.\ref{fig:rank_3s_baryon3} we plot the diagram for the
fragmentation of our chain into a nonleading antibaryon--baryon pair.
The antibaryon and baryon do not contain any quarks from the original
hadrons involved in the collision. Therefore, each of the quark or
antiquarks involved can be strange, the antibaryon and the baryon can be
double--strange or even triple--strange hyperons, for instance $\bar
\Xi$, $\Xi$, $\bar \Omega$ or $\Omega$ hyperons. But we note according
to this mechanism the fragmentation into double--strange or
triple--strange anti--hyperons is favored against the fragmentation into
double--strange or
  triple--strange hyperons. This diagram (or slight variations of it) 
  is the only diagram available
  in the PYTHIA/JETSET chain fragmentation to produce
  antihyperon--hyperon pairs in  diquark fragmentation. This explains
  the wong behaviour of PYTHIA and DPMJET-III discussed above.  

In order to correct the behaviour of DPMJET-III we have to add
one missing diagram in diquark fragmentation. We did not 
correct\footnote{This has the advantage, that we continue to use the
standard PYTHIA and that we therefore will be able to replace PYTHIA
with new versions.
}
PYTHIA, but we added the missing fragmentation step im DPMJET--III before
calling PYTHIA. 

The missing diagram: We note, in the PYTHIA chain fragmentation
diquark--antidiquark pairs can be exchanged in any position  
(see Fig.\ref{fig:rank_3s_baryon3}) except near to the chain--end
diquarks. In Fig.\ref{fig:rank_2s_baryon3} we plot the missing diagram.
At the chain--end we obtain a diquark--antidiquark pair, this has
to fragment into a pair of mesons. Next to this in the chain we obtain a
baryon. All three quarks of his baryon can be strange quarks.
Therefore, in this diagram we can obtain double--strange and
triple--strange hyperons, which dominate against the anti--hyperons
produced eventually later in the chain fragmentation. Obviously, this is
the missing mechanism needed in DPMJET-III. 

In order to introduce the mechanism according to
Fig.\ref{fig:rank_2s_baryon3} we have to introduce a new parameter,
which describes the probability, that the new diagram is to be used in
the first diquark fragmentation step. We obtain good results with
the rather small probability of such a baryonium  $B_{2 Mesons}$ = 0.01. 
For the flavor distribution of this baryonium and the exchanged
diquark--antidiquark pair we use a rather small value of $r_s=5 \% $ .

\begin{figure}[h]
\begin{center}
\includegraphics[height=8cm,width=8cm]{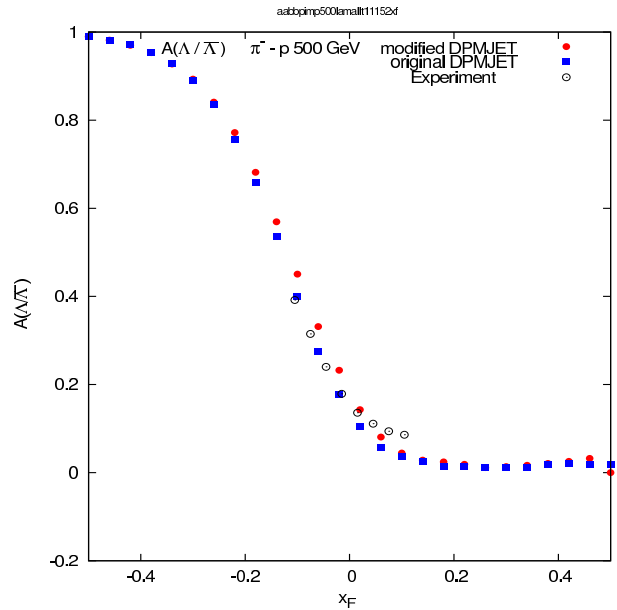}
\end{center}

\caption{ 
The $A({\Lambda}/{\bar \Lambda})$ asymmety. Plotted are the original
DPMJET-III, the modified DPMJET-III and the experimental data from the
E791 Collaboration \cite{Aitala2000}.
 }

\vspace{4mm}
\label{fig:aabbpimp500lamallt11152xf}
\end{figure}

In Fig.\ref{fig:aabbpimp500lamallt11152xf} we present the result for the
 $A({\Lambda}/{\bar \Lambda})$ asymmety. We do not expect, that the
 mechanism according to
Fig.\ref{fig:rank_2s_baryon3} changes the $\Lambda$ or $\bar \Lambda$
production in an essential way. Indeed we find in
Fig.\ref{fig:aabbpimp500lamallt11152xf} that the original DPMJET-III and
the changed model agree perfectly with each other as well as with the
data from the E791 Collaboration \cite{Aitala2000}.

\begin{figure}[h]
\begin{center}
\includegraphics[height=8cm,width=8cm]{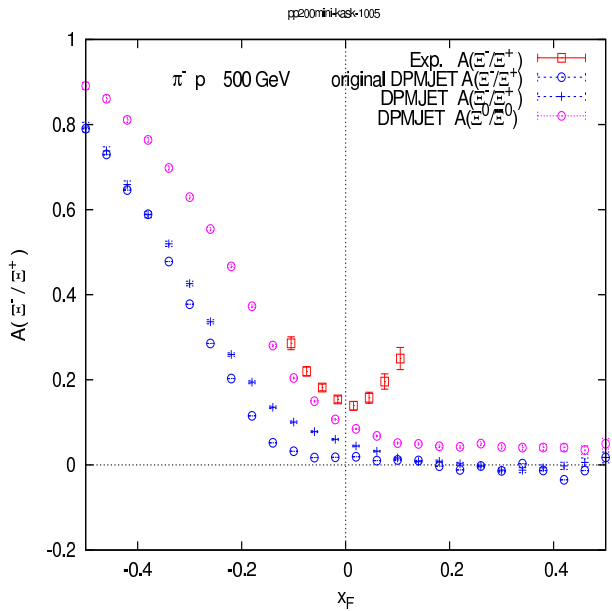}
\end{center}

\caption{ 
The $A({\Xi}/{\bar \Xi})$ asymmety. Plotted are the original
DPMJET-III, the modified DPMJET-III and the experimental data from the
E791 Collaboration \cite{Aitala2000}.
 }

\vspace{-1mm}
\label{fig:aabbpimp500kasct11192xf}
\end{figure}

\begin{figure}[h]
\begin{center}
\includegraphics[height=8cm,width=8cm]{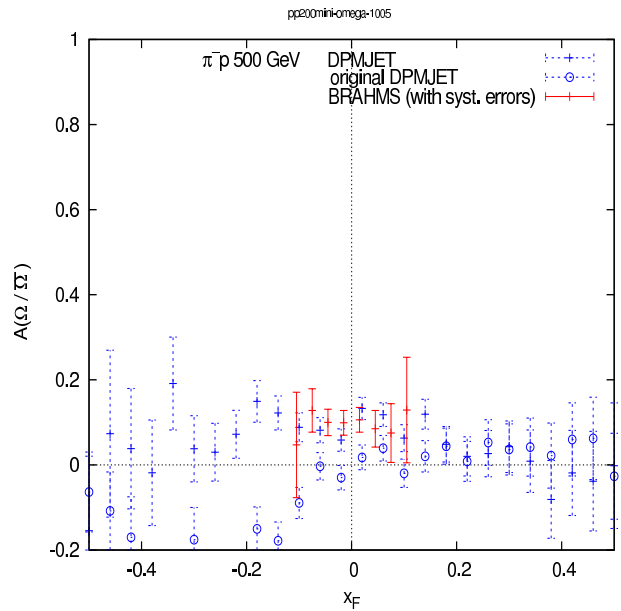}
\end{center}

\caption{ 
The $A({\Omega}/{\bar \Omega})$ asymmety. Plotted are the original
DPMJET-III, the modified DPMJET-III and the experimental data from the
E791 Collaboration \cite{Aitala2000}.
 }

\vspace{-1mm}
\label{fig:aabbpimp500omegat11192xf}
\end{figure}

 The $A({\Xi}/{\bar \Xi})$ and $A({\Omega}/{\bar \Omega})$ 
 asymmeties are significantly modified by the 
 mechanism according to
Fig.\ref{fig:rank_2s_baryon3}. We plot both asymmetries in 
Figs.\ref{fig:aabbpimp500kasct11192xf} and
\ref{fig:aabbpimp500omegat11192xf}. Again the asymmeties according to
the original and modified DPMJET-III are compared to the 
data from the E791 Collaboration \cite{Aitala2000}. With the original
DPMJET-III the asymmety $A({\Xi}/{\bar \Xi})$ is zero in the central
region and the asymmetry  $A({\Omega}/{\bar \Omega})$ is even
negative in the central region. With the modified DPMJET-III both
asymmetries become positive in the central region like the experimental
data.
For the choosen strangeness probability the $\Omega$-asymmetry is somewhat too
strong while the $\Xi$-asymmtry is not sufficient.

\begin{figure}[h]
\begin{center}   
\includegraphics[height=8cm,width=8cm]{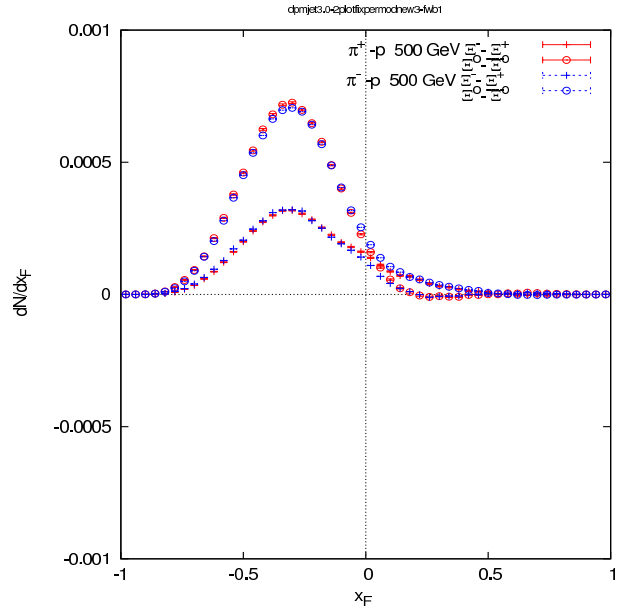}
\end{center}

\caption{ The net $\Xi - \bar{\Xi}$ contribution for charged an neutral
hyperons for $\pi^+ p$ and $\pi^- p$ scattering.
 }
  
\vspace{-1mm}
\label{fig:KasK_collect}
\end{figure}

Most significant seems the rise of of the $\Xi$ asymmetry in the forward region. To investigate 
the influence of the meson isospin   the net hyperon contributions are 
considered for various isospin combinations  in  Fig.\ref{fig:KasK_collect}. 
The direct influence $\pi^{-}\rightarrow\cdots d\rightarrow dss$ seems not to reach the
central region where baryons are sufficently significant. Seen is a rank 2 effect $\pi^{-}\rightarrow\cdots\bar{u}\rightarrow\bar{u}d+\bar{d}\bar{s}\bar{s}$
which balancing contribution $\pi^{-}\rightarrow\cdots d\rightarrow\bar{d}d+\bar{d}\bar{s}\bar{s}$
is suppressed as $m(\eta)\gg m(\pi^0)$. 

In conlusion the $\pi^-$ isospin effect is not only too small to
explain the effect but it actually contributes in the wrong direction. 

The errorbars are significant If the rise is confirmed by a second experiment again a new effect
will have to added. It could evidence the backward peak postulated
in \cite{bopp2000} caused by a tiny three chain forward-backward exchange contribution
leaving a pion and possibly a $ssd$ on the opposite side.

 Rapidity distributions of  $p$, $\bar p$, $\Lambda$,
$\bar \Lambda$, $\Xi$ and $\bar \Xi$ were measured in 
proton--proton collisions at 158 GeV by  
 the NA49 Collaboration \cite{NA49pp158}. Also in this experiment the
 $\bar \Xi$/$\Xi$ ratio in the central region is found to be
 0.44$\pm$0.08 that is smaller then 1. In the original DPMJET-III as
 well as in PYTHIA this ratio is found to be approximately equal to 1.
 The reasons for this are exactly the same as discussed above. In order
 to get a better agreement to the data, we have to modify DPMJET-III in
 the same way as described above, that is we have to include the
 mechanism according to Fig.\ref{fig:rank_2s_baryon3}. 

\begin{figure}[h]
\begin{center}
\includegraphics[height=8cm,width=8cm]{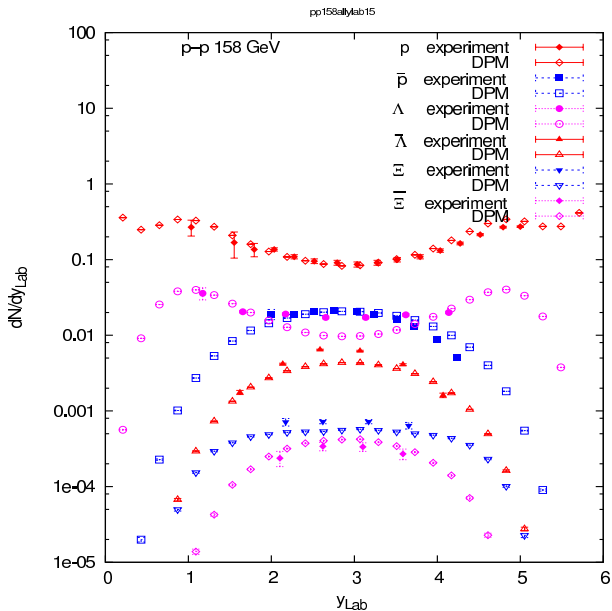}
\end{center}

\caption{ Rapidity distributions of produced $p$, $\bar p$, $\Lambda$,
$\bar \Lambda$, $\Xi$ and $\bar \Xi$ in proton--proton collisions at 158
GeV. We compare the rapidity distributions according to the modified
DPMJET-III with experimental data from the NA49 Collaboration
\cite{NA49pp158}.
 }

\vspace{-1mm}
\label{fig:pp158allylab15}
\end{figure}

 In Fig.\ref{fig:pp158allylab15} we compare the results of the modified
 DPMJET-III (using $B_{2 Mesons}$ = 0.02 und $r_s=30\% $) with the data from
 the NA49 Collaboration \cite{NA49pp158}. We find excellent agreements
 for $p$, $\bar p$, $\Lambda $ and $\bar \Lambda $ production and improved
 agreements for $\Xi$ and $\bar \Xi$ production.

\section{ Summary }
 
 Experimental data  are extremely useful to
 improve hadronic production models like {\sc Dpmjet}-III.
 Of particular importance in this 
 respect are data on hadron production in hadron--hadron
 collisions and d-Au collisions measured at RHIC. In 
 these collisions we found partly already in  previous 
papers\cite{Ranft20001,Ranft20004,Ranft20003} 
  three important
 corrections to be applied to {\sc Dpmjet}--III: 
\begin{description}
\item(i)~~Percolation and
 fusion of chains -  the data from RHIC allow to determine the amount of
 percolation to be implemented into {\sc Dpmjet}--III.\vspace*{-2mm} ,
 see \cite{Ranft20001,Ranft20003}.
\item(ii) ~Collision
 scaling of large $p_{\perp}$ hadron production in d--Au 
 collisions, see \cite{posterpap04}.\vspace*{-2mm} 
\item(iii) Replacing the Gaussian
 transverse momentum distribution contained in the JETSET--Pythia code
\cite{JETSET,AND85}   
  by an exponential distribution in soft hadronic
 collisions, see \cite{Bopp2004}.\vspace*{-2mm}
 \item(iv)~Implementation of new diagrams for an improved description of
 baryon stopping, see Figs. \ref{diqsea} and \ref{new2}.
 \vspace*{-2mm}
 \item(v)~~Adding the mechanism 
  according to Fig.\ref{fig:rank_2s_baryon3} to the fragmentation of
  diquark chains
\end{description}   
These corrections are somewhat technical, they do not involve the basic 
structure of the string model.  In the considered not too dense region
(hadron--hadron collisions and nuclear collisions involving light
nuclei)
the general features
of the model can be considered as quite reliable.
 
 In string models the baryon quantum number 
 are carried by vortex lines which
 play a special role in the string structure. 
 The understanding of baryon-  and, in 
 particular, the net-baryon-production is therefore of 
 central importance.

 In DPMJET--III, baryon stopping diagrams
 have to be included to get a consistent description of the considered
RHIC data.  With the new RHIC data, 
supporting anomalous baryon stopping,
this is no longer a merely theoretical exercise. 
 Good agreement with the critical experimental data was obtained.

\vfill
 
 \bibliographystyle{prsty}

 \bibliography{dpm11}

\begin{thebibliography}{10}

\bibitem{Capella94a}
A.~Capella, U.~Sukhatme, C.~I.~Tan, and J.~Tr\^an Thanh~V\^an, Phys.\ Rep.\
  {\bf 236},  227  (1994).

\bibitem{Engel95a}
R. Engel, Z.\ Phys.\ {\bf C66},  203  (1995).

\bibitem{Engel95d}
R. Engel and J. Ranft, Phys.\ Rev.\ {\bf D54},  4244  (1996).

\bibitem{Roesler20001}
S. Roesler, R. Engel, and J. Ranft, {Proceedings of ICRC 2001, Copernicus Ges.} 
{\bf {1}},   (2001).

\bibitem{Roesler20002}
S. Roesler, R. Engel, and J. Ranft, { [hep-ph/hep--ph/0012252], Proc. of Monte Carlo
  2000, Lisboa, Oct.2000,Springer,p.1033 } {\bf {}},  {1033}  (2000).

\bibitem{posterpap04}
{F.W.Bopp, J.Ranft, R.Engel and S.Roesler},  [hep--ph/0403084 (unpublished).

\bibitem{NA35FIN}
{NA35 Collaboration, T.~Alber {\it et al.} }, Z. Phys. {\bf C 64},  195  (1994).

\bibitem{Alber98}
{NA35 Collaboration, T.~Alber {\it et al.} }, Eur. Z. Phys. {\bf C2},  643  (1998).


\bibitem{Aitala2000}
{E.M.Aitala {\it et al.} E791 Collaboration }, Phys.\ Lett.\ B {\bf {\bf 469}},  {9}
  (2000 [hep-ex/0009016]).

\bibitem{NA49pp158}
{T.Susa {\it et al.} NA49 Collaboration}, Nucl. Phys. A {\bf {\bf 698}},  {491c}
  (2002).


\bibitem{Kharzeev96}
D. Kharzeev, Phys.Lett. {\bf B 378},  238  (1996).

\bibitem{Capella96}
A. Capella and B. Kopeliovich, Phys. Lett. {\bf B381},  325  (1996).

\bibitem{Capella99a}
A. Capella, E.~G. Ferreiro, and C.~A. Salgado, Phys.\ Lett.\ {\bf B459},  27
  (1999).

\bibitem{Capella99}
A. Capella and C.~A. Salgado,  Phys.\ Rev.\  C {\bf 60} (1999) 054906 {[hep-ph/9903414]}.

\bibitem{Vance99}
S.~E. Vance and M.~Gyulassy, Phys.\ Rev.\ Lett. {\bf {83}},  {1735}  (1999).

\bibitem{Casado99}
J.~A. Casado, Nucl.Phys. {\bf {A 651}},  {93}  (1999).

\bibitem{Ranft20001}
J. Ranft, [hep--ph/0002137] (unpublished).

\bibitem{Ranft20004}
J. Ranft, R. Engel, and S. Roesler, {Proceedings of ICRC 2001, Copernicus Ges.
  } {\bf {}},  {435}  (2001).

\bibitem{Ranft20003}
J. Ranft, R. Engel, and S. Roesler, { [hep--ph/0012112], Proc. of Monte Carlo
  2000, Lisboa, Oct.2000,Springer, } {\bf {}},  {979}  (2000).

\bibitem{Braun97}
M.~A. Braun, C. Pajares, and J. Ranft, Int. J. Mod. Phys. {\bf A 14},  2689
  (1999).

\bibitem{Braun2000a}
M. Braun and C. Pajares, Eur. Phys. J. {\bf {C16}},  {359}  (2000).

\bibitem{Ranft2003}
{J.Ranft, R.Engel and S.Roesler}, Nucl. Phys. B (Proc. Suppl.) {\bf {122}},
  {292}  (2003).

\bibitem{Bopp2004}
{F.W.Bopp, J.Ranft, R.Engel and S.Roesler}, [hep--ph/410027] (to be published in proceedings).

\bibitem{BackPHOBOS03}
{B.B.Back et al, PHOBOS Colaboration }, Phys.\ Rev.\  C {\bf 70} (2004) 011901 {[nucl-ex/0309013]}.

\bibitem{SteinbergPHOBOS04}
{P.Steinberg {\it et al.}, PHOBOS Collaboration, J. Phys. G;Nucl. Part. Phys.. {\bf
  {30}},  {S683}  (2004)}ls
.

\bibitem{STAR0412}
{F.Simon, STAR Colaboration }, Acta Phys.\ Hung.\  A {\bf 27} (2006) 287 [nucl-ex/0412031] .

\bibitem{BRAHMS0401}
{I.~G.~Bearden  {\it et al.}, BRAHMS Collaboration}, Phys.\ Rev.\ Lett.\ {\bf 94}, 032301 (2005) , [nucl-ex/0401025].

\bibitem{BRAHMS0409}
{I.~G.~Bearden  {\it et al.}, BRAHMS Collaboration}, Phys.\ Lett.\  B {\bf 607} (2005) 42, [nucl-ex/0409002].

\bibitem{Rossi97}
G.~C. Rossi and G. Veneziano, Nucl. Phys. {\bf B123},  507  (1977).

\bibitem{SUKHA}
{ U.~Sukhatme, K.~Lassila and R.~Orava}, Phys.Rev. {\bf D25},  2075  (1982).

\bibitem{KOPEL}
{ B.~Z.~Kopeliovich and B.~G.~Zakharov}, Z. Phys. {\bf C43},  241  (1989).

\bibitem{BAMJET1}
J. Ranft and S. Ritter, Acta \ Phys. \ Pol. \ B {\bf 11},  259  (1980).

\bibitem{BAMJET}
S. Ritter, Comput. \ Phys. \ Commun. {\bf 31},  393  (1984).

\bibitem{JETSET}
T. Sj\"ostrand, Comp. Phys. Comm. {\bf 82},  74  (1994).

\bibitem{AND85}
{ B.~Andersson, G.~Gustafson and T.~Sj\"ostrand}, Physica Scripta {\bf 32},
  574  (1985).

\bibitem{Capella99b}
{A~Capella, E.G.~Ferreiro, and C.A.~Salgado}, Nucl.\ Phys.\ {\bf {A 661}},
  {502}  (1999).

\bibitem{Boppshabelski}
{F.W.Bopp and Yu.M.Shabelski},   Phys.\ Atom.\ Nucl.\  {\bf 68}, 2093 (2005)
  [hep--ph/0406158]; {F.W.Bopp and Yu.M.Shabelski},   
  Eur.\ Phys.\ J.\  A {\bf 28}, 237 (2006) [hep--ph/0603193].

\bibitem{auren77}
{P.~Aurenche and F.~W.~Bopp}, Nucl.\ Phys.\ {\bf {B119}},  {}  (1977).

\bibitem{bopp2000}
{F.~W.~Bopp},  [hep--ph/0002190] (unpublished).

\bibitem{BBC75}
{Alper, B. {\it at al.}}, Nucl.\ Phys.\ {\bf B100},  237  (1975).


\bibitem{Aguilar91a}
M. Aguilar-Benitez {\it et~al.}, Z.\ Phys.\ {\bf C50},  405  (1991).

\bibitem{Arakelian:2002iw}
{ G.~H.~Arakelian, A.~Capella, A.~B.~Kaidalov and Y.~M.~Shabelski}, Eur.\
  Phys.\ J.\ C  {\bf 26},  {81}  (2002).

\bibitem{STARptopip}
{J.C.Dunlop for the STAR Collaboration}, Nucl.\ Phys.\ A774:139-148, 2006 [nucl--ex/0510073].

\bibitem{AdamsSTAR06}
J.~Adams {\it et al.},  STAR Collaboration,
Phys.\ Lett.\  B {\bf 637} (2006) 161
  [nucl-ex/0601033].



\bibitem{Capella:2002sx}
{A.~Capella}, Phys.\ Lett.\ B  {\bf 542},  {65}  (2002).

\bibitem{Liu2003}
{F.M.Liu {\it et al.}}, Phys. Rev. D {\bf 67},  {034011}  (2003).

\end{thebibliography}
 

\end{document}